# Structure and magnetism of S = 1/2 kagome antiferromagnets $NiCu_3(OH)_6Cl_2$ and $CoCu_3(OH)_6Cl_2$

**Yue-sheng Li[1] and Qing-ming Zhang[1,2]**

[1]Department of Physics, Renmin University of China, Beijing 100872, P. R. China.

[2]Corresponding author. Tel. /Fax.: +86-10-62517767. E-mail address: qmzhang@ruc.edu.cn

**Abstract.** We have successfully synthesized S = 1/2 kagome antiferromagnets $MCu_3(OH)_6Cl_2$ (M = Ni and Co) by hydrothermal method with rotating pressure vessel. Structural characterization shows that both compounds have the similar crystal structure as $ZnCu_3(OH)_6Cl_2$ with R-3m symmetry. Similar to $ZnCu_3(OH)_6Cl_2$, the compounds show no obvious hysteresis at 2 K. A spin glass transition is found in both $NiCu_3(OH)_6Cl_2$ and $CoCu_3(OH)_6Cl_2$ at low temperatures (6.0 and 3.5 K respectively) by AC susceptibility measurements. It indicates no long-range magnetic order and a strong spin frustration. The substitution of $Zn^{2+}$ by magnetic ions $Ni^{2+}$ or $Co^{2+}$, effectively enhances interlayer exchange coupling and changes the ground state of the kagome spin system.

**KEYWORDS:** spin frustration, kagome antiferromagnet, spin glass

## 1. Introduction

In a spin frustrated system, a spin cannot find a proper orientation to favor all the interactions with its neighboring spins. The frustration is caused either by competing interactions or by spin geometric configuration with antiferromagnetic (AF) nearest-neighbor interaction. Many interesting and unexpected consequences caused by spin frustration, are not well understood so far [1]. Among spin frustrated systems, low-dimensional $S = 1/2$ antiferromagnet with highly geometric frustration has attracted particular attention [2]. For this kind of antiferromagnets, novel quantum states such as resonating-valence-bond (RVB) and "spin-liquid" ground state have been proposed. The novel concepts are considered to have an intimate connection with high-temperature superconductivity in cuprates [3]. Structurally perfect $S = 1/2$ kagome compounds are rare. $S = 1/2$ kagome antiferromagnet $ZnCu_3(OH)_6Cl_2$ was successfully synthesized, which has been regarded as the first "structure-perfect" $S = 1/2$ frustrated compound in some aspects [4-6]. Recently synthesized β-Vesignieite $BaCu_3V_2O_8(OH)_2$ was reported to be structurally perfect and exhibits long-range order at 9 K due to Dzyaloshinki-Moriya (DM) interaction [7]. In $ZnCu_3(OH)_6Cl_2$, kagome layers are separated by non-magnetic and non-Jahn-Teller-active $Zn^{2+}$ ions, which predominantly occupy interlayer triangular sites. Structurally perfect kagome layers with R-3m space group are determined to be preserved even down to 2 K [8]. No magnetic transition was observed at least down to 50 mK and the magnetism was considered as quantum paramagnetic state at low temperature [9,10], which is in accord with many theoretical results using approximations and numerical simulations. On the other hand, there are many debates on the occupancy of Zn in kagome planes in $ZnCu_3(OH)_6Cl_2$ [11-14]. Non-magnetic defects in kagome planes can partially release the geometric frustration while $Cu^{2+}$ ions entering into interlayer triangular sites bring interlayer magnetic coupling. These will substantially affect spin ground state. The question is what we can expect for the kagome system with a stronger interlayer magnetic coupling.

In this paper, we performed a comparative study on the structures and magnetic properties of $ZnCu_3(OH)_6Cl_2$ and herbertsmithite-like kagome antiferrromagnets $MCu_3(OH)_6Cl_2$ (M = Ni and Co). $Ni^{2+}$ and $Co^{2+}$ are non-Jahn-Teller-active magnetic ions with $S = 1$ and $3/2$ respectively and prefer interlayer sites [4]. It is expected that the two magnetic ions may introduce novel interactions and spin states which may open a new window for exploring kagome antiferromagnets. In fact, several groups have tried to synthesize $NiCu_3(OH)_6Cl_2$ and $CoCu_3(OH)_6Cl_2$ [15-17]. Unfortunately the attempts with a conventional static hydrothermal method failed or CuO impurities cannot be removed completely.

By employing hydrothermal method with rotating pressure vessel, so-called dynamic hydrothermal method, we have successfully synthesized $NiCu_3(OH)_6Cl_2$ and $CoCu_3(OH)_6Cl_2$. X-ray diffraction patterns and Rietveld refinements show that the compounds are well single-phased without any sign of CuO and other impurities. Structural characterizations reveal that the crystal structures of both compounds are similar to that of $ZnCu_3(OH)_6Cl_2$. The interlayer triangular sites are dominantly occupied by $M^{2+}$ ions and kagome sites by $Cu^{2+}$ ions (figure 1). As the bond angle $\angle MOCu$ is close to the critical value of $95^o$ [18,19] (M=Ni, Co), the exchange interaction between interlayer $Ni^{2+}/Co^{2+}$ and nearest-neighbor $Cu^{2+}$ in kagome planes falls in the critical region of AF and ferromagnetic (FM) coupling and is much smaller than in-plane Cu-Cu AF exchange interactions. In this sense, the compounds $NiCu_3(OH)_6Cl_2$ and $CoCu_3(OH)_6Cl_2$ still preserve the essences of 2D kagome antiferromagnets in both structure and magnetism, and will play an important and unique role in exploring the effects of spin frustration.

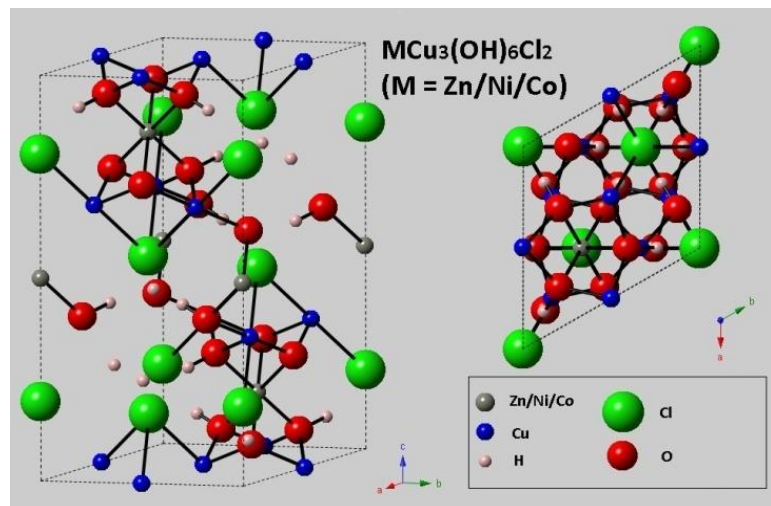

**Figure 1.** The crystal structure of $MCu_3(OH)_6Cl_2$ (M = Zn, Ni or Co).

## 2. Materials and methods

ZnCu$_3$(OH)$_6$Cl$_2$, NiCu$_3$(OH)$_6$Cl$_2$ and CoCu$_3$(OH)$_6$Cl$_2$ powder samples were synthesized under dynamic and hydrothermal conditions provided by a homogeneous reactor. A 30 mL teflon liner was charged with 498 mg of CuCO$_3$ Cu(OH)$_2$ xH$_2$O(4.5 mmol Cu), 10 mL of water, and 307 mg (**1:** 2.25 mmol Zn) of ZnCl$_2$, 535 mg (**2:** 2.25 mmol Ni) NiCl$_2$ 6H$_2$O, or 535 mg (**3:** 2.25 mmol Co) CoCl$_2$ 6H$_2$O. The liner was capped and placed into a stainless steel pressure vessel. The pressure vessel was fixed to the ratation axis of a homogeneous reactor. The rotation rate of pressure vessel was fixed to be 35 r/min all the time. The vessel was heated to 205 °C at a rate of 1 °C/min, and its temperature was maintained for 48 hr. and then cooled down to room temperature at a rate of 0.1 °C/min. A blue-green polycrystalline powder was found at the bottom of each vessel, isolated from the liner by filtration, washed with water, and dried over by a loft drier (at 70 °C). By the procedure, 610 mg of ZnCu$_3$(OH)$_6$Cl$_2$, 590 mg of NiCu$_3$(OH)$_6$Cl$_2$ and 610 mg of CoCu$_3$(OH)$_6$Cl$_2$ were obtianed, which correspond to the yields of 95%, 93% and 96% respectively, with respect to the starting material CuCO$_3$ Cu(OH)$_2$ xH$_2$O. The product does not include any agglomerate pieces, which suggests a homogenous and complete reaction process. We compared ZnCu$_3$(OH)$_6$Cl$_2$ samples synthesized by dynamic and static method. They have a very close quality. Magnetization measurements demonstarte that the sample synthesized by dynamic method has a slightly lower occupancy of Cu at interlayer sites than that by static method (~ 17% vs. ~ 21%). X-ray diffraction measurements were carried on a Shimadzu XRD-7000 diffractometer using Cu K$_\alpha$ radiation ($\lambda$ = 1.5403 Å). The diffraction data were processed and fit using Rietveld techniques with the GSAS program [20] using the same crystal structure model as ZnCu$_3$(OH)$_6$Cl$_2$ (R-3m) [4]. Inductively coupled plasma (ICP) measurements were made with a HORIBA Jobin Yvon Ultima 2 ICP system. The measured ratio of M:Cu is very close to 1:3 in all the three samples (1.03(4):2.97 in ZnCu$_3$(OH)$_6$Cl$_2$; 1.02(4):2.98 in NiCu$_3$(OH)$_6$Cl$_2$; 1.06(4):2.94 in CoCu$_3$(OH)$_6$Cl$_2$). Magnetic measurements were made with a Physical Property Measurement System (PPMS) by Quantum Design.

### 3. Results and discussion

X-ray powder diffraction patterns and Rietveld refinements are shown in figure 2. No additional peak is seen, which implies that CuO and other impurity phases are negligible in the samples. The refinement results suggest that the crystal grains in powder samples are well crystallized. The grain size is estimated to be ~200 nm for the three samples from full width at half maximum (FWHM) [20]. Due to the chemical similarity among $Zn^{2+}$, $Ni^{2+}$ and $Co^{2+}$, the structural parameters are very close (table 1), as well as the site exchange. This is confirmed by very similar diffraction patterns, relative intensities and FWHMs of the three samples. The crystal symmetry of R-3m is kept in both $NiCu_3(OH)_6Cl_2$ and $CoCu_3(OH)_6Cl_2$, indicating the occupation fraction of Cu at triangular sites is smaller than a critical value. For example, the symmetry change from rhombohedral (R-3m, x > 0.33) to monoclinic (P 21/n, x < 0.33) was observed in $Zn_xCu_{4-x}(OH)_6Cl_2$ [4].

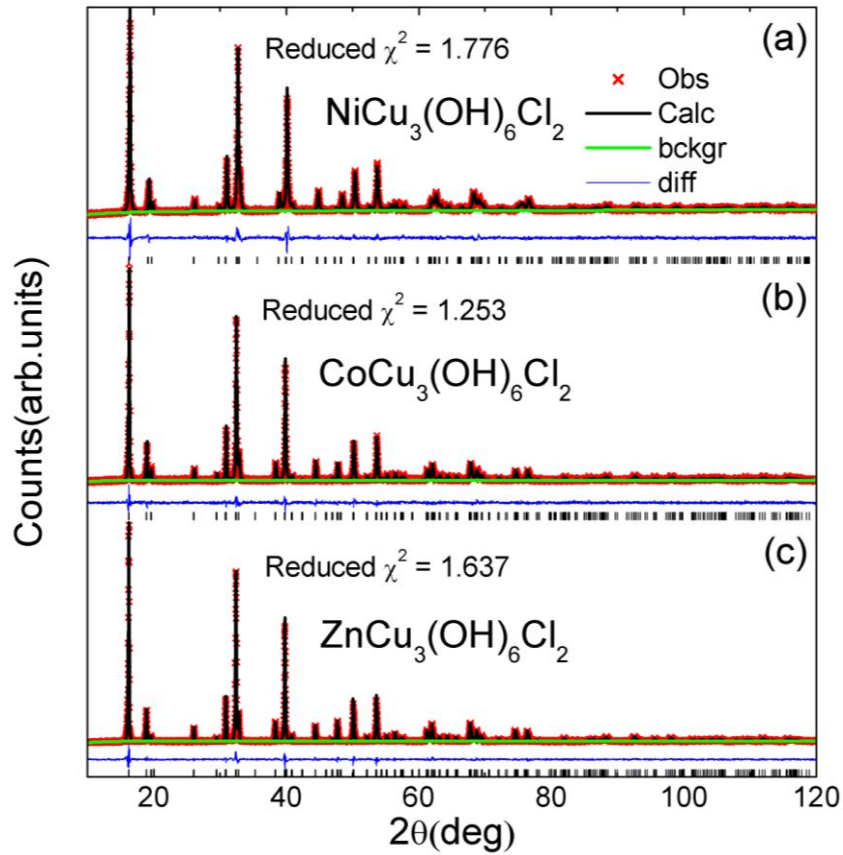

**Figure 2.** Powder X-ray diffraction Rietveld refinements in (a), (b) and (c) for $NiCu_3(OH)_6Cl_2$, $CoCu_3(OH)_6Cl_2$ and $ZnCu_3(OH)_6Cl_2$ respectively at T = 295 K.

Table 1. X-ray refinement results[a]

| MCu$_3$(OH)$_6$Cl$_2$ (R -3 m) | | M = Ni | M = Co | M = Zn |
|---|---|---|---|---|
| a (Å) | | 6.8505(8) | 6.8416(6) | 6.8418(4) |
| c (Å) | | 13.9288(18) | 14.0934(14) | 14.0954(8) |
| M | Uiso*100 | 0.36(11) | 0.57(11) | 1.77(7) |
| Cu | Uiso*100 | 1.46(7) | 1.79(7) | 1.56(5) |
| O | x = - y | 0.20421(30) | 0.20578(30) | 0.20542(23) |
|  | z | 0.06475(25) | 0.06286(26) | 0.06066(20) |
|  | Uiso*100 | 1.26(19) | 1.27(19) | 1.35(15) |
| Cl | z | 0.19411(21) | 0.19448(22) | 0.19513(17) |
|  | Uiso*100 | 0.78(12) | 1.43(12) | 1.70(9) |
| M-O (Å) | | 2.0887(33) | 2.1035(33) | 2.1285(26) |
| Cu-O (Å) | | 2.0104(20) | 1.9963(20) | 1.9839(15) |
| Cu-Cl (Å) | | 2.7697(21) | 2.7803(22) | 2.7741(17) |
| ∠Cu-O-Cu (deg) | | 116.83(19) | 117.92(19) | 119.12(15) |
| ∠M-O-Cu (deg) | | 96.12(11) | 96.90(11) | 96.48(9) |

[a]Considering the insignificant scattering contribution of proton, the positional parameters of proton are fixed during the refinements according to Ref [4].

The temperature dependence of magnetization (or DC susceptibility) from 2 to 300 K with an applied field of 2000 Oe, is shown in figure 3 (a). All the magnetization curves exhibit a Curie-like tail at low temperatures (7 K ~ 100 K), which are attributed to the magnetic ions (or defects) at interlayer sites [21,22]. CoCu$_3$(OH)$_6$Cl$_2$ shows the largest magnetization in the whole temperature range. And the magnetization of NiCu$_3$(OH)$_6$Cl$_2$ and ZnCu$_3$(OH)$_6$Cl$_2$ decreases gradually. This can be understood through spin moments of Co$^{2+}$ (S = 3/2) and Ni$^{2+}$ (S = 1). The inset of figure 3(a) shows the Curie-Weiss fitting at low temperatures, the fitting constants are c ~ 0.520(2) Kcm$^3$/mol Cu; θ ~ -12.17(2) K and c ~ 0.991(3) Kcm$^3$/mol Cu; θ ~ -9.19(4) K for NiCu$_3$(OH)$_6$Cl$_2$ and CoCu$_3$(OH)$_6$Cl$_2$, respectively. Supposing a full occupation of interlayer sites by Ni or Co, the Curie constant c gives effective spin moment μ$_{eff}$ ~ 3.5μ$_B$ for Ni$^{2+}$ and 4.9μ$_B$ for Co$^{2+}$. The effective moments are consistent with the reported ones [23]. Both the Curie constant c and the Weiss temperature θ are much larger than those of ZnCu$_3$(OH)$_6$Cl$_2$ (c ~ 0.0287(1) Kcm$^3$/mol Cu; θ ~ -1.08(1) K) [21]. This suggests a much stronger interlayer coupling in NiCu$_3$(OH)$_6$Cl$_2$ and CoCu$_3$(OH)$_6$Cl$_2$. The interlayer coupling will substantially affect magnetism at low temperatures (T < |θ|) in the two compounds. Figure 3(b) shows Curie-Weiss fit at high temperatures (200 ~ 300 K). The Weiss temperatures given by high-T mean field

fitting exhibit a monotonic evolution with effective moments of interlayer ions. And the Weiss temperatures of $NiCu_3(OH)_6Cl_2$ (-100 K) and $CoCu_3(OH)_6Cl_2$ (- 40 K) are much lower than that of $ZnCu_3(OH)_6Cl_2$ (-380 K). The high-temperature susceptibilities contain the contributions from both interlayer and kagome ions. So the Weiss temperatures are no longer a good measure for interlayer or intralayer coupling. Interestingly, the monotonic decrease of the Weiss temperatures was also observed in $Zn_xCu_{4-x}(OH)_6Cl_2$ and $Mg_xCu_{4-x}(OH)_6Cl_2$ when increasing the interlayer magnetic ions $Cu^{2+}$ [5, 24]. The Weiss temperatures given by high-temperature susceptibilities may be a rough indicator of paramagnetic-like magnetizations of interlayer magnetic ions. High-T fitting also gives the Curie constants of 0.79, 0.87 and 1.28 Kcm$^3$/mol Cu for $ZnCu_3(OH)_6Cl_2$, $NiCu_3(OH)_6Cl_2$ and $CoCu_3(OH)_6Cl_2$, respectively. Approximately they are consistent with spin quantum numbers, g-factors, and the composition of the Cu, Ni and Co ions. As mentioned above, high-temperature susceptibilities are contributed by both interlayer and kagome magnetic ions. It is difficult to separate the total bulk susceptibilities into the two parts. So a precise consistency is limited.

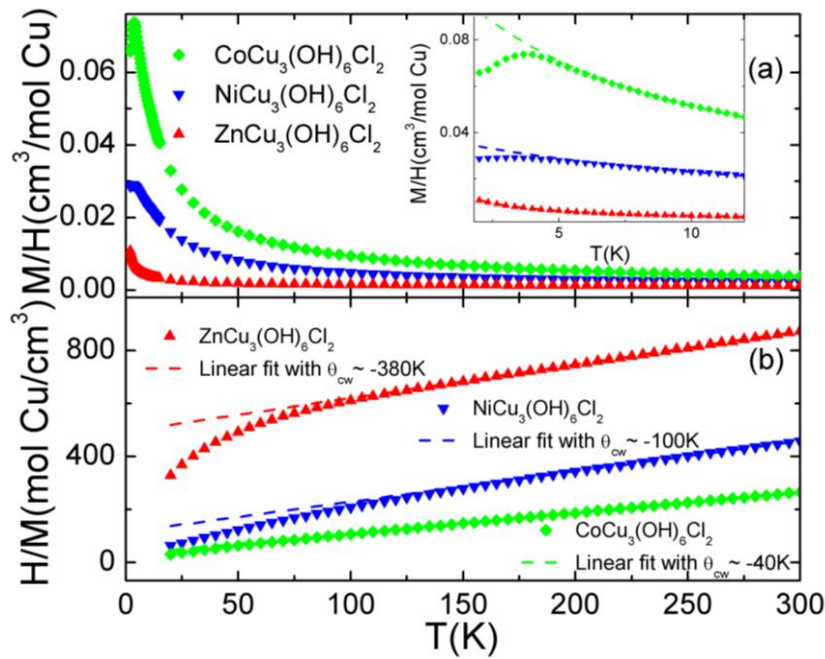

**Figure 3.** (a) Temperature dependence of magnetization under an applied field of 2000 Oe for the three samples. The inset shows the data in the low temperature range (2 ~ 13 K) and the dashed lines are extrapolated from Curie-like tails above magnetic freezing temperature (T > $T_s$). (b) Fitting high-temperature data (200 ~ 300 K) to Curie-Weiss law.

Most interestingly, a deviation from Curie-Weiss law ("kink") is clearly observed in the inset of figure 3(a) in $MCu_3(OH)_6Cl_2$ (M = Ni and Co). In order to get deeper insight into this kink, we further compare susceptibilities under ZFC and FC from 2 to 10 K, as shown in figure 4. For $ZnCu_3(OH)_6Cl_2$, there is no difference between FC and ZFC in the whole temperature range, which is consistent with the reported results and reflects a quantum paramagnetic state [6]. While an obvious splitting develops below a characteristic temperature $T_s \sim 6.0$ K and 3.5 K for $NiCu_3(OH)_6Cl_2$ and $CoCu_3(OH)_6Cl_2$, respectively. These exactly correspond to the kinks in figure 3(a). Surprisingly, the transition temperatures vary little compared with Clinoatacamite ($Cu_2(OH)_3Cl$) [5, 26, 27]. We notice that clinoatacamite $Cu_2(OH)_3Cl$, $NiCu_3(OH)_6Cl_2$ and $CoCu_3(OH)_6Cl_2$, have almost identical ∠MOCu angles (M refers to interlayer magnetic ions), which are very close to the critical angle of 95 degrees. We think this may cause an interlayer coupling at the same level of ~10 K in the three compounds and hence similar transition temperatures. For $Cu_2(OH)_3Cl$ it is more complicated because the slight distortions of kagome planes by interlayer Jahn-Teller-active $Cu^{2+}$ ions need to be taken into account. While for non-Jahn-Teller-active $Ni^{2+}$ and $Co^{2+}$, perfect kagome planes are preserved. So it can be safely said that the transition is driven by interlayer magnetic coupling considering that no magnetic transtion was observed in herbertsmithite above 2 K. Moreover, one can see that the kink temperature under ZFC is a little higher than that under FC in $CoCu_3(OH)_6Cl_2$, as shown in figure 4. It is a hint for spin-glass behaviour and we will come back to this point later.

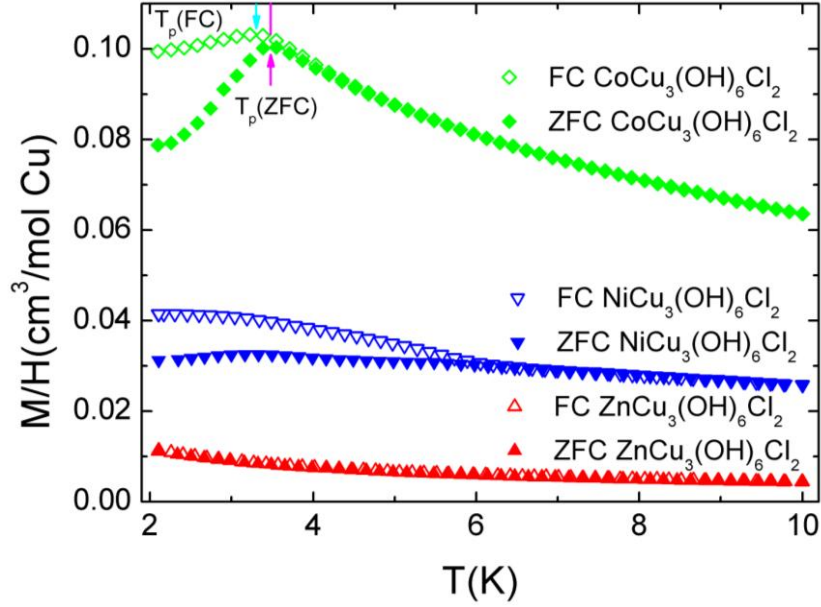

**Figure 4.** (a) Temperature dependence of DC susceptibility under zero-field-cooling (ZFC) and field-cooling (FC) condition ($H_{cooling} = H_{measure} = 100$ Oe). The colored arrows mark the kink/peak temperatures ($T_p(FC)$ and $T_p(ZFC)$) for $CoCu_3(OH)_6Cl_2$.

M-H measurements are shown in figure 5. No obvious hysteresis can be seen at 2 K in all the samples, which excludes the possibility that ferromagnetic components exist in the compounds. This is clearly different from the Mg- and Zn-compounds with smaller x [17,24,25], in which a $Cu_2(OD)_3Cl$-like distortion around interlayer $Cu^{2+}$ may develop at low temperatures due to Jahn-Teller effect. And it is reported that a ferromagnetic transition occurs at ∼ 6 K in $Cu_2(OD)_3Cl$ [26,27,28]. These confirm that interlayer sites are dominantly occupied by non-Jahn-Teller-active magnetic ions $M^{2+}$ (M = Zn, Ni and Co) in these three samples.

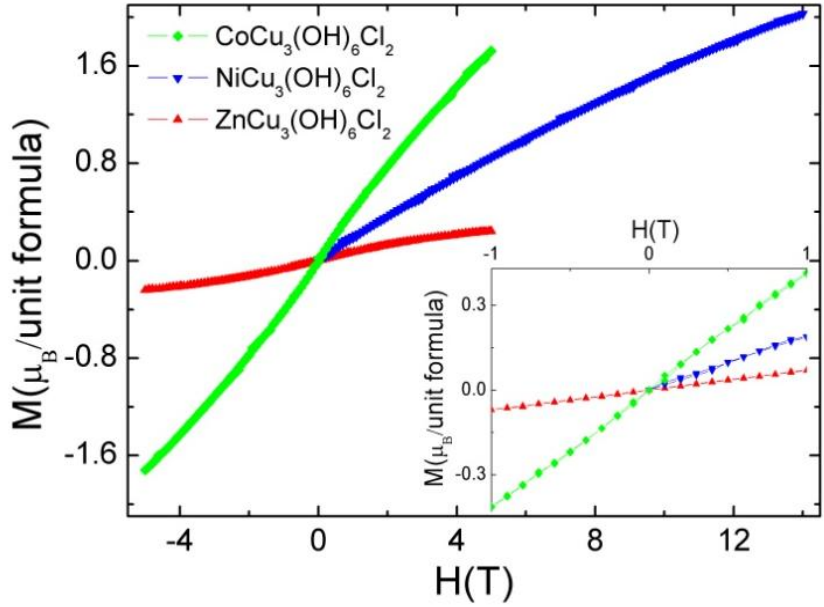

**Figure 5.** Magnetic field dependence of magnetization measured at 2 K. For $ZnCu_3(OH)_6Cl_2$ and $CoCu_3(OH)_6Cl_2$, intact loops were measured at a constant rate of 100 Oe/sec. For $NiCu_3(OH)_6Cl_2$, the applied field was first increased to 14 T, then decreased to 0 T at a constant rate of 50 Oe/sec.

The low-temperature kink/splitting is a prominent feature of the Ni- and Co-compounds. The most possible origin is AF ordering or spin-glass transition. In figure 6, we present AC susceptibilities in $CoCu_3(OH)_6Cl_2$ and $NiCu_3(OH)_6Cl_2$. For $CoCu_3(OH)_6Cl_2$, the kink/splitting position shifts to a higher temperature with increasing measurement frequencies, which is a typical spin-glass behaviour and rule out the possibility of AF ordering. The shift in $NiCu_3(OH)_6Cl_2$ is a little obscure but still visible.

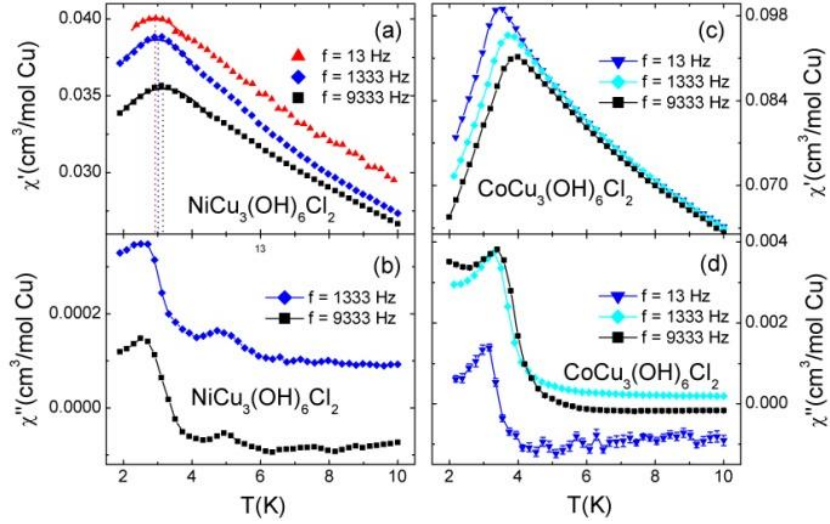

**Figure 6.** AC susceptibilities are measured under zero applied field. (a) and (b) The real and imaginary parts of AC susceptibilities of $NiCu_3(OH)_6Cl_2$. The dashed lines mark the corresponding peak temperatures, 2.9, 3.0 and 3.2 K for 13 Hz, 1333Hz and 9333 Hz respectively; (c) and (d) The real and imaginary parts of AC susceptibilities of $CoCu_3(OH)_6Cl_2$.

At present we do not exactly know which kind of spins participate in spin-glass transition: interlayer or in-plane or even both spins? However it is certain that spin-glass state in $NiCu_3(OH)_6Cl_2$ and $CoCu_3(OH)_6Cl_2$ reflects a subtle change of spin configuration in kagome layers, which is ultimately connected with interlayer magnetic spins in some way. It is plausible that the spin-glass transition may occur on interlayer triangular sites. But a direct interaction between interlayer spins seems less possible because the distance between interlayer spins is very large and there is no anion to bridge the exchange interaction between them. The chemical disorder may not be the driving force of spin-glass transition in this spin system. In fact, spin-glass transition has been reported in $S = 5/2$ kagome antiferromagnet hydronium jarosite $(H_3O)Fe_3(SO_4)_2(OH)_6$ [29], in which it was argued that planar anisotropy rather than chemical disorder drives spin-glass transition [30]. Further theoretical and experimental efforts are highly required to understand the novel spin-glass state in $NiCu_3(OH)_6Cl_2$ and $CoCu_3(OH)_6Cl_2$.

**4. Conclusions**

In conclusion, we have successfully synthesized single-phased kagome compounds $NiCu_3(OH)_6Cl_2$ and $CoCu_3(OH)_6Cl_2$. X-ray diffraction refinements reveal that the compounds have a crystal structure similar to that of $ZnCu_3(OH)_6Cl_2$. Magnetic measurements show that a kink or FC/ZFC splitting appears at several Kelvins in $NiCu_3(OH)_6Cl_2$ and $CoCu_3(OH)_6Cl_2$. No obvious magnetization hysteresis loop accompanies the kink. Further AC susceptibility measurements demonstrate that the kink position shifts to higher temperatures with increasing measurement frequencies, which is a typical spin-glass behaviour. On one hand it means no long-range magnetic order develops down to 2 K, which suggests that $NiCu_3(OH)_6Cl_2$ and $CoCu_3(OH)_6Cl_2$ are still good candidates for studying strong spin frustration. On the other hand, the substitution of Zn by magnetic ions effectively enhances interlayer coupling in the kagome systems and brings out new physics. The kagome compounds and the novel spin-glass state in them will inspire further experimental and theoretical efforts in the future.

**Acknowledgements**

This work was supported by the 973 program under Grant Nos. 2011CBA00112&2012CB921701, by the NSF of China under Grant No. 11034012, by the Fundamental Research Funds for Central Universities and the Research Funds of RUC.